\pdfoutput=1

\documentclass[11pt]{article}

\usepackage{longtable} 
\usepackage{array} 
\usepackage{caption} 

\usepackage[dvipsnames]{xcolor}
\usepackage{authblk}
\usepackage{graphicx}
\usepackage{xurl}
\usepackage{enumitem}
\usepackage{xspace}
\usepackage{xparse}
\usepackage{float}

\DeclareDocumentCommand\newstep{o}{%
\item\IfNoValueTF{#1}{}{#1 \textendash\xspace}}
\newlist{steps}{enumerate}{1}
\setlist[steps]{label=\textit{RQ \arabic*:},leftmargin=*}

\usepackage[preprint]{acl}

\usepackage{times}
\usepackage{latexsym}
\usepackage{colortbl}
\usepackage{amssymb}
\usepackage{tablefootnote}
\usepackage{amsmath}
\usepackage{hyperref}
\usepackage{cleveref}
\crefrangelabelformat{section}{#3#1#4--#5\crefstripprefix{#1}{#2}#6}
\usepackage{subcaption}
\usepackage{dblfloatfix}

\Crefname{section}{Appendix}{Appendices}

\usepackage{ragged2e}
\usepackage{tcolorbox}

\usepackage{listings}
\definecolor{codegreen}{rgb}{0,0.6,0}
\definecolor{codegray}{rgb}{0.5,0.5,0.5}
\definecolor{codepurple}{rgb}{0.58,0,0.82}
\definecolor{backcolour}{rgb}{0.95,0.95,0.92}

\lstdefinestyle{mystyle}{
    backgroundcolor=\color{backcolour},   
    commentstyle=\color{codegreen},
    keywordstyle=\color{magenta},
    numberstyle=\tiny\color{codegray},
    stringstyle=\color{codepurple},
    basicstyle=\ttfamily\footnotesize,
    breakatwhitespace=false,         
    breaklines=true,                 
    captionpos=b,                    
    keepspaces=true,                 
    numbers=left,                    
    numbersep=5pt,                  
    showspaces=false,                
    showstringspaces=false,
    showtabs=false,                  
    tabsize=2
}

\usepackage[T1]{fontenc}

\usepackage[utf8]{inputenc}

\usepackage{booktabs}
\usepackage{multirow}
\usepackage{makecell} 
\usepackage{arydshln}

\usepackage{tikz}

\usepackage{microtype}

\usepackage{cuted}

\usepackage{inconsolata}

\newcommand{\dream}{Dream\xspace}
\newcommand{\llada}{LLaDA\xspace}

\newcommand{\llava}{LLaVA-1.6\xspace}
\newcommand{\qwen}{Qwen2.5-VL\xspace}
\newcommand{\mmada}{MMaDA\xspace}
\newcommand{\lavida}{LaViDa\xspace}
\newcommand{\auto}{Autoregressive VLMs\xspace}
\newcommand{\diff}{Diffusion VLMs\xspace}

\newcommand{\eg}{\hbox{\emph{e.g.,}}\xspace}
\newcommand{\ie}{\hbox{\emph{i.e.,}}\xspace}

\newcolumntype{L}[1]{>{\raggedright\let\newline\\\arraybackslash\hspace{0pt}}m{#1}}
\newcolumntype{C}[1]{>{\centering\let\newline\\\arraybackslash\hspace{0pt}}m{#1}}
\newcolumntype{R}[1]{>{\raggedleft\let\newline\\\arraybackslash\hspace{0pt}}m{#1}}

\usepackage{tabularx}

\definecolor{lightgreen}{RGB}{230,255,230}
\definecolor{lightblue}{RGB}{230,230,255}
\definecolor{lightred}{RGB}{255,230,230}
\definecolor{lightgrey}{RGB}{245,245,245}

\definecolor{YaleYellow}{RGB}{179, 176, 4} 
\definecolor{NYUPurple}{RGB}{134, 1, 175}  
\definecolor{NTUBlue}{RGB}{2,2,200} 
\definecolor{Alibaba}{RGB}{255, 106, 0}
\definecolor{Center}{RGB}{0, 128, 0}
\definecolor{UCASRed}{RGB}{150, 0, 0}      

\title{Analyzing Diffusion and Autoregressive Vision Language Models\\in Multimodal Embedding Space}

\author{
\bf{
Zihang Wang$^{*\hspace{.1em}\textcolor{NYUPurple}{\boldsymbol{S}}}$
\quad
Siyue Zhang$^{*\hspace{0.02em}{\textcolor{NTUBlue}{\boldsymbol{N}}}}$$^{\hspace{.02em}{\textcolor{Alibaba}{\boldsymbol{A}}}}$ 
\quad
Yilun Zhao$^{\hspace{.1em}\textcolor{YaleYellow}{\boldsymbol{Y}}}$ 
\vspace{-9pt}}
\quad
Jingyi Yang$^{\hspace{0.02em}{\textcolor{NTUBlue}{\boldsymbol{N}}}}$$^{\hspace{.02em}{\textcolor{Alibaba}{\boldsymbol{A}}}}$ 
\\
\bf{
Tingyu Song$^{\hspace{.1em}\textcolor{UCASRed}{\boldsymbol{U}}}$
\quad
Anh Tuan Luu$^{\hspace{.1em}\textcolor{NTUBlue}{\boldsymbol{N}}}$ \quad
Chen Zhao$^{\hspace{.1em}\textcolor{NYUPurple}{\boldsymbol{S}}}$
$^{\hspace{.1em}\textcolor{Center}{\boldsymbol{C}}}$
}
\vspace{9pt}\\
$^{\textcolor{NTUBlue}{\boldsymbol{N}}}$Nanyang Technological University \quad
$^{\textcolor{YaleYellow}{\boldsymbol{Y}}}$Yale University \quad
$^{\textcolor{NYUPurple}{\boldsymbol{S}}}$NYU Shanghai \quad
\\
$^{\textcolor{Alibaba}{\boldsymbol{A}}}$Alibaba-NTU Singapore Joint Research Institute 
\\
$^{\textcolor{UCASRed}{\boldsymbol{U}}}$University of the Chinese Academy of Sciences
\\
$^{\textcolor{Center}{\boldsymbol{C}}}$Center for Data Science, New York University

}

\begin{document}

\maketitle

\begin{abstract}

Embedding models are a fundamental component of modern AI systems such as semantic search and retrieval-augmented generation. Recent advances in large foundation models have substantially accelerated the development of embedding models, including those based on Large Language Models (LLMs), Vision Language Models (VLMs), and Multimodal LLMs. More recently, Large Diffusion Language Models (dLLMs) and Multimodal dLLMs have emerged as competitive alternatives to autoregressive models, offering advantages such as bidirectional attention and parallel generation. This progress naturally raises a critical yet unexplored question: can Multimodal dLLMs serve as effective multimodal embedding models? To answer this, we present the first systematic study of converting Multimodal dLLMs into embedding models. We evaluate state-of-the-art Multimodal dLLMs and Autoregressive VLMs across three categories of embedding tasks: classification, visual question answering, and information retrieval. Our results show that Multimodal dLLM embeddings generally underperform their autoregressive VLM counterparts. The stronger diffusion-based model, \lavida, lags by only 3.5 points on classification, 2.5 points on VQA, and 4.4 points on retrieval tasks, whereas the other diffusion-based model, \mmada, exhibits substantially larger performance gaps, exceeding 20 points across all tasks. Further analysis reveals insufficient image–text alignment in diffusion-based models, accounting for the observed limitations in their embedding performance.\footnote{Code and results are available at \url{https://github.com/ZihangWang93/diff_vlm2vec}.}

\end{abstract}

\section{Introduction}

Embedding models map textual, visual, and audio inputs into fixed-dimensional vector representations (\ie embeddings). These vectors encode the semantic content of the input, enabling efficient comparison, retrieval, and ranking, and serving as a fundamental component of modern AI systems such as semantic search \citep{dpr,bge_gemma,gme,lco} and retrieval-augmented generation (RAG) \citep{rag1,rag_paper,mrag,browsecomp}. 

\begin{figure}[t!]
    \centering
    \includegraphics[width=0.48\textwidth]{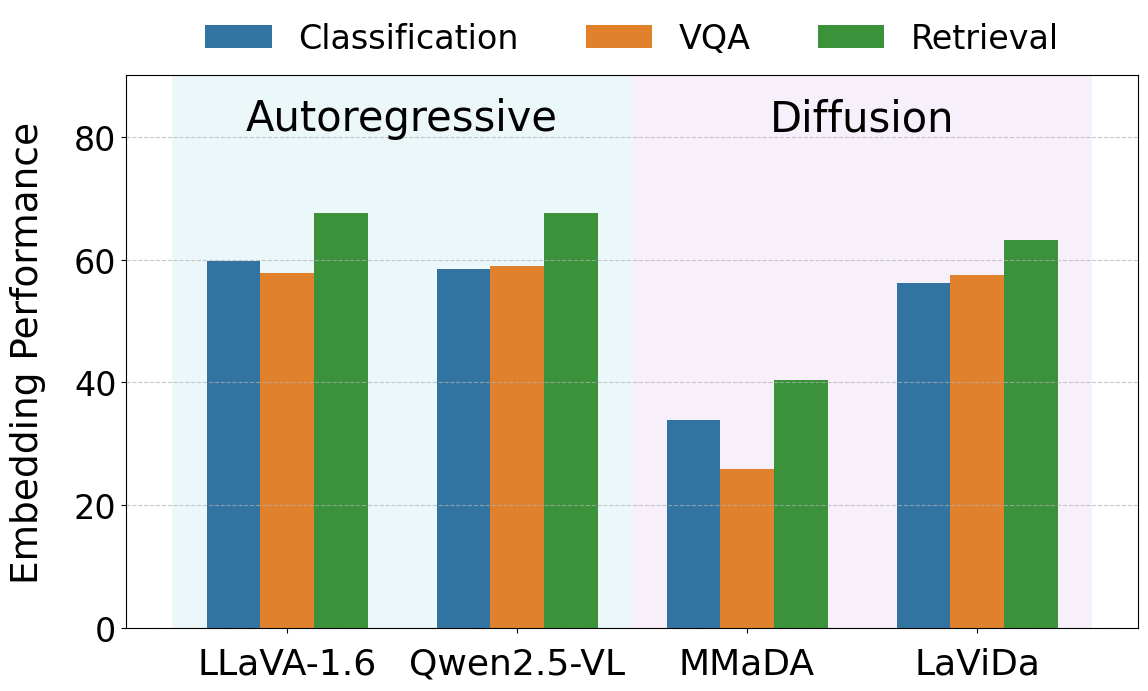}
    \captionsetup{justification=justified, singlelinecheck=false}  
    \caption{Average performance on three multimodal embedding meta-tasks. Overall, Diffusion VLM embeddings underperform Autoregressive VLM embeddings, despite the use of bidirectional attention; however, performance varies substantially across diffusion models, with \lavida remaining competitive while \mmada shows a substantial gap.}
    \label{fig1}
\end{figure}

Recent progress in large foundation models has substantially accelerated the development of embedding models. Advances in Large Language Models (LLMs) have given rise to state-of-the-art text embedding models such as E5 \citep{e5}, BGE \citep{bge_gemma}, NV-Embed \citep{nv_embed}, and Qwen3 Embedding \citep{qwen3embed}. In parallel, progress in Vision Language Models (VLMs) has enabled multimodal embedding models including GME \citep{gme}, Jina Embeddings \citep{jina_v4}, and RzenEmbed \citep{rzen}. To further support additional modalities such as audio and video, recent work adapts Multimodal Large Language Models (MLLMs) into omni-embedding models, exemplified by Omni-Embed-Nemotron \citep{omni_nemotron} and LCO-Embedding \citep{lco}.
Despite these successes, large foundation models may exhibit limitations when applied to embedding tasks. In particular, LLM-based embedding models often struggle to capture global context due to the causal attention structure adopted during large-scale pretraining \citep{bellm,echo,diffembed}.

Large Diffusion Language Models (dLLMs) have recently emerged as a compelling alternative to autoregressive LLMs, attracting growing attention. Recent dLLMs formulate language modeling as a discrete diffusion process via forward token masking and reverse unmasking under a bidirectional attention architecture, \eg \llada \citep{llada} and \dream \citep{dream}. With billions of model parameters and large-scale pretraining, these models demonstrate performance competitive with state-of-the-art autoregressive LLMs such as Qwen2.5 \citep{qwen} and LLaMA3 \citep{llama3} across a range of tasks. More importantly, DiffEmbed \citep{diffembed} shows that the diffusion-based embedding model, benefiting from inherent bidirectional attention, is particularly effective at modeling global context, yielding superior representations for long and structurally complex documents.

Building on these advances, recent work has extended dLLMs to the multimodal setting, resulting in Multimodal Diffusion Language Models\footnote{For brevity, we refer to Multimodal Diffusion Language Models as \diff, and to Vision Language Models with autoregressive LM backbones as \auto.}. In practice, \diff, \eg \llada-V \citep{llada_v} and \lavida \citep{lavida}, incorporate a vision encoder and MLP connector that projects visual features into the language embedding space, enabling effective multimodal alignment. In contrast, \mmada \citep{mmada} adopts an image tokenizer and a unified diffusion architecture over both text and image tokens. While \llada-V and \lavida are trained solely for multimodal understanding, \mmada is additionally optimized for text-to-image generation. Despite these differences, these \diff have achieved strong performance on multimodal understanding and reasoning benchmarks \citep{hudson2019gqa,mmmu,mme}. Such progress naturally raises an unexplored question: \textit{can Diffusion VLMs serve as effective multimodal embedding models, and how do they compare to autoregressive counterparts across tasks?}

\begin{figure*}[t!]
    \centering
    \includegraphics[width=0.95\textwidth]{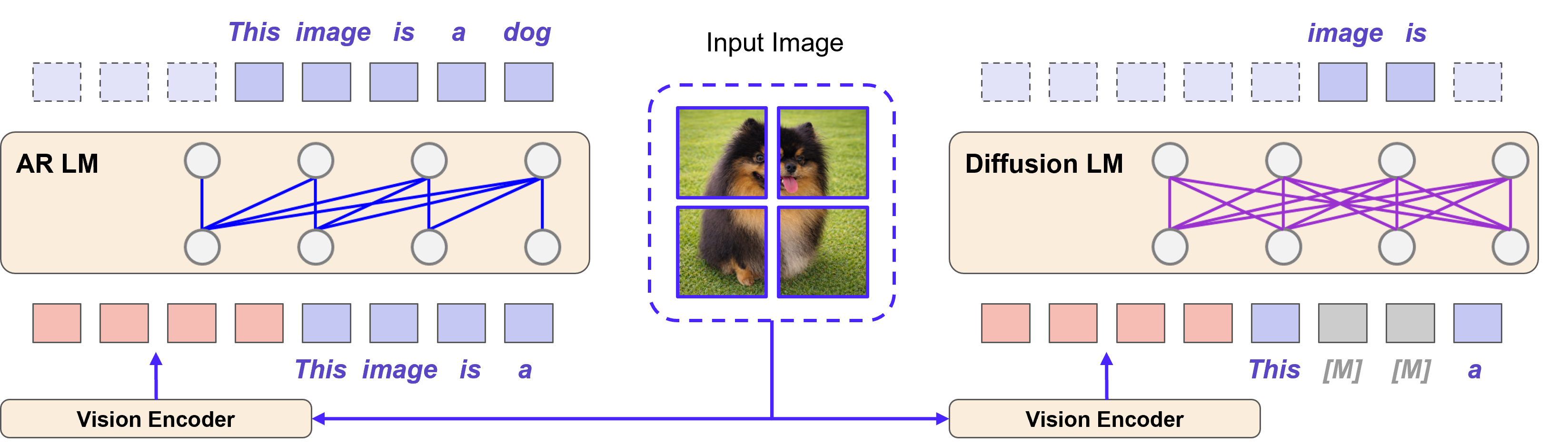}
    \captionsetup{justification=justified, singlelinecheck=false}  
\caption{General architectures of \auto (left) and \diff (right). For both types of models, the input image is encoded into tokens by a vision encoder (\eg SigLIP \citep{siglip2} and VQ-GAN \citep{vqgan}) and interleaved with text tokens as input to the language model. \auto are pretrained using causal attention for next-token prediction, while \diff are pretrained using bidirectional attention for token unmasking.}

    \label{fig2}
\end{figure*}

In this work, we present the first systematic evaluation of Diffusion and Autoregressive VLM embedding models. We select two diffusion models (\ie \mmada and \lavida) and two autoregressive models (\ie \llava and \qwen) as baselines, representing state-of-the-art performance. Following VLM2Vec \citep{vlm2vec}, we apply contrastive fine-tuning to these pretrained models and evaluate both in-domain and out-of-domain embedding performance across 32 datasets spanning three meta-tasks: classification, visual question answering, and information retrieval.

Our experimental results show that \diff models generally underperform \auto models across diverse tasks, despite their use of bidirectional attention. Notably, \lavida and \mmada exhibit markedly different behaviors. \lavida shows a relatively modest gap compared to \auto, with average drops of 3.5 points on classification, 2.5 points on VQA, and 4.4 points on retrieval tasks. In contrast, \mmada demonstrates a substantially larger performance degradation, exceeding 20 points across all meta-tasks. Further analysis indicates that Autoregressive VLM embedding models achieve significantly higher performance with limited fine-tuning data than diffusion-based counterparts, supporting our finding that image–text representations in \diff are insufficiently aligned.

To summarize, our contributions include:

\begin{itemize}[leftmargin=*]

\item We introduce Diffusion VLM embedding models, a new class of embedding models derived from Multimodal Large Diffusion Language Models.
 
\item We provide the first systematic comparative analysis of Diffusion and Autoregressive VLM embedding models.

\item We reveal that Diffusion VLM embeddings generally underperform autoregressive models despite bidirectional attention; however, \lavida exhibits only a modest performance gap and strong out-of-domain generalization.

\end{itemize}

\section{Background}
\label{sec: beckground}

\paragraph{Multimodal Embedding Tasks.} Multimodal embedding tasks aim to learn a unified representation space in which semantically related content from different modalities, most commonly text and images, can be directly compared and retrieved. Early efforts focused on restricted settings such as single-modal retrieval, where both the query and target belong to the same modality \citep{fu2023nights}, and cross-modal retrieval, where the query and target are drawn from different modalities \citep{wang2021n24news}. More recent formulations generalize these tasks toward an any-to-any paradigm, allowing both queries and candidates to consist of arbitrary modality combinations, including text, images, audio, or their compositions \citep{circo,ma2024wikissnq,mrmr}. Under this unified perspective, a broader range of multimodal problems, such as classification, visual question answering, information retrieval, and visual grounding, have been framed as embedding-based ranking tasks \citep{mieb,vlm2vec}.

\paragraph{Multimodal Embedding Models.} To map inputs from different modalities into a shared latent space, early approaches primarily adopted contrastive pretraining frameworks that align paired cross-modal data, as exemplified by CLIP-style models trained on large-scale image–text pairs \citep{clip,evaclip,siglip2}. Another line of work employs modality-specific encoders to process different modal inputs independently, followed by projection or fusion into a common embedding space \citep{uniir,vista}. More recent approaches fine-tune pretrained VLMs for embedding tasks using either text-only \citep{e5v} or multimodal supervision \citep{vlm2vec}, typically deriving the final representation from the last transformer layer via pooling or last-token selection. To further extend modality coverage, recent efforts adapt MLLMs into omni-embedding models, enabling unified representation learning across additional modalities such as audio and video \citep{omni_nemotron,sailembedding}. Notably, most prior methods are built upon large pretrained models with autoregressive LLM backbones, \eg GME \citep{gme} and VLM2Vec \citep{vlm2vec}.

\begin{table*}[t!]
\centering
\renewcommand{\arraystretch}{1}
\captionsetup{justification=justified,singlelinecheck=false}
\begin{tabularx}{\textwidth}{
    >{\raggedright\arraybackslash}l 
    >{\centering\arraybackslash}X
    >{\centering\arraybackslash}X
    >{\centering\arraybackslash}X
    >{\centering\arraybackslash}X
}
\toprule
\textbf{Dataset} 
& \textbf{\llava}
& \textbf{\qwen}
& \textbf{\mmada}
& \textbf{\lavida} \\
\midrule
ImageNet-1K        & 68.9 & \textbf{72.7} & 27.0 & 64.3 \\
N24News            & \textbf{81.1} & 76.6 & 62.6 & 71.3 \\
HatefulMemes       & \textbf{70.1} & 67.2 & 49.9 & 50.6 \\
VOC2007            & \textbf{92.9} & 85.4 & 66.4 & 84.9 \\
SUN397             & \textbf{71.6} & 68.9 & 45.0 & 68.2 \\

\rowcolor{black!6}
Place365           & \textbf{39.7} & 38.9 & 30.4 & 35.2 \\
\rowcolor{black!6}
ImageNet-A         & 39.2 & \textbf{45.5} & 2.4  & 42.0 \\
\rowcolor{black!6}
ImageNet-R         & 77.4 & 78.5 & 34.5 & \textbf{79.3} \\
\rowcolor{black!6}
ObjectNet          & 44.2 & 39.4 & \textbf{62.6} & 57.2 \\
\rowcolor{black!6}
Country-211        & \textbf{12.0} & 11.2 & 3.9  & 8.8 \\
\midrule
\textit{All Classification} & \textbf{59.7} & 58.4 & 33.9 & 56.2 \\

\bottomrule
\end{tabularx}
\caption{Performance comparison across 10 classification tasks. Best results are shown in bold for each task. Tasks highlighted in gray are in-domain, meaning their data have been included during contrastive fine-tuning.}
\label{tab:classification_results}
\end{table*}

\paragraph{Multimodal Large Diffusion Language Models.} Unlike autoregressive LLMs, dLLMs model the data distribution by reconstructing corrupted token sequences via an iterative denoising process. Recent methods such as \llada \citep{llada} and \dream \citep{dream} propose discrete diffusion formulations for token sequences, in which a forward process progressively masks tokens and a reverse process jointly predicts all masked positions using bidirectional context. The models are trained with a cross-entropy loss applied only to masked tokens. With scaling to billions of model parameters and training on large-scale data, these masked diffusion models have demonstrated strong performance on challenging math and code reasoning tasks \citep{dream}.

Building on this paradigm, recent studies extend dLLMs to the multimodal setting by incorporating vision encoders and fine-tuning on multimodal corpora, yielding Multimodal Large Diffusion Language Models, namely \diff, \eg LLaDA-V \citep{llada_v}, \lavida \citep{lavida}, and \mmada \citep{mmada}. Uniquely, \mmada unifies multimodal understanding and text-to-image generation within a single diffusion framework. More recently, DiffusionVL \citep{diffusionvl} adapts pretrained autoregressive VLMs to the diffusion paradigm, rather than building upon pretrained diffusion language models.\footnote{As DiffusionVL was released after this submission, we will evaluate it in the further experiment.} However, the effectiveness of multimodal embeddings produced by these diffusion models has yet to be explored. In this work, we take the first step to fill this gap.

\section{Diffusion VLM Embeddings}
\label{sec: method}

In this work, we focus on a new class of multimodal embedding models built upon \diff. In line with Autoregressive VLM embedding models, Diffusion VLM embedding models integrate visual inputs via a vision encoder and processes images and text jointly within the backbone VLMs. The final embedding is derived from the contextualized token representations in the last transformer layer. Following \citet{vlm2vec}, Autoregressive VLM embeddings are obtained using the last-token representation, whereas Diffusion VLM embeddings are aggregated via mean pooling, following \citet{diffembed}. To learn effective multimodal representations, we fine-tune Diffusion VLMs using a contrastive learning objective, consistent with the training strategy adopted for autoregressive embedding models.

During contrastive fine-tuning, both the query $q$, comprising image tokens, an instruction, and text content, and the multimodal target $t$ are encoded into dense vectors $\mathbf{h}$  \citep{vlm2vec}. We optimize the standard InfoNCE loss $\mathcal{L}$ with positive targets $t^+$ and negatives $t^-$, including both in-batch and hard negatives:
\begin{equation}
\resizebox{0.8\hsize}{!}{$
\min \; \mathcal{L}
= - \log
\frac{
\phi\!\left( \mathbf{h}_{q}, \mathbf{h}_{t^{+}} \right)
}{
\phi\!\left( \mathbf{h}_{q}, \mathbf{h}_{t^{+}} \right)
+ \sum_{t^{-}}
\phi\!\left( \mathbf{h}_{q}, \mathbf{h}_{t^{-}} \right)
}
$,}
\label{eqn:infonce}
\end{equation}
where $\phi(\mathbf{h}_q, \mathbf{h}_t)$ is the temperature-scaled cosine similarity function between query $q$ and target $t$.

The primary distinction between Autoregressive and Diffusion VLM embedding models lies in their attention architectures, as illustrated in \Cref{fig2}. Autoregressive models are pretrained using causal attention and a next-token prediction objective, which restricts each token to attend only to its preceding context. In contrast, diffusion models employ a denoising objective that reconstructs corrupted inputs, allowing tokens to be processed jointly \citep{diffembed}. During contrastive fine-tuning, both paradigms retain their pretraining attention patterns and differ only in the optimization objective: autoregressive embedding models continue to operate under causal attention, whereas diffusion-based models natively support bidirectional attention across tokens.

\section{Experimental Setups}
\label{experimental_setups}

To compare Diffusion and Autoregressive VLM embeddings, we fine-tune a range of pretrained models on multiple embedding tasks. In this section, we introduce the evaluation tasks (\S\ref{tasks}), representative models (\S\ref{models}), and implementation details (\S\ref{impl}).

\begin{table*}[t!]
\centering
\renewcommand{\arraystretch}{1}
\captionsetup{justification=justified,singlelinecheck=false}
\begin{tabularx}{\textwidth}{
    >{\raggedright\arraybackslash}l 
    >{\centering\arraybackslash}X
    >{\centering\arraybackslash}X
    >{\centering\arraybackslash}X
    >{\centering\arraybackslash}X
}
\toprule
\textbf{Dataset} 
& \textbf{\llava}
& \textbf{\qwen}
& \textbf{\mmada}
& \textbf{\lavida} \\
\midrule

OK-VQA             &   \textbf{73.1} &   58.1  & 42.0 &  61.1 \\
A-OKVQA            &   \textbf{59.6} &   50.8   & 30.4 &   51.6 \\
DocVQA             &   78.7 &   \textbf{90.1}   & 17.2 &  82.9 \\
InfographicsVQA    &   42.2 &   \textbf{72.6}   & 17.8 &   48.6 \\
ChartQA            &   48.7 &   \textbf{63.0}  & 14.3 &  51.1 \\
Visual7w           &   51.2 &  51.5   & 38.8 &   \textbf{54.3} \\
\rowcolor{black!6}
ScienceQA          &   \textbf{40.1} &   39.9  & 18.9 &   39.2 \\
\rowcolor{black!6}
VizWiz             &   \textbf{47.7} &   43.3   & 12.9 &   43.3 \\
\rowcolor{black!6}
GQA                &   60.7 &   47.4   & 51.1 &   \textbf{67.0} \\
\rowcolor{black!6}
TextVQA            &   \textbf{76.1} &   72.9   & 15.2 &   75.8 \\
\midrule
\textit{All VQA}   &  57.8 &   \textbf{59.0}   & 25.9 &   57.5 \\

\bottomrule
\end{tabularx}
\caption{Performance comparison across 10 VQA tasks. Best results are shown in bold for each task. Tasks highlighted in yellow are in-domain, meaning their data have been included during contrastive fine-tuning.}
\label{tab:vqa_results}
\end{table*}

\subsection{Tasks}
\label{tasks}

Following \citet{vlm2vec}, we adopt three meta-tasks—classification, VQA, and retrieval—to evaluate multimodal embedding models. These tasks are formulated as ranking problems, where both the query and candidate documents are embedded into a shared latent space. The ground-truth label, answer, or gold document is expected to achieve the highest similarity score with the query.


\paragraph{Classification.} In this meta-task, the query consists of an instruction and an image, optionally accompanied by related text, while the target corresponds to a class label. The candidate set contains all possible class labels, with the number of candidates ranging from binary classification to 1,000 classes. These tasks require the embedding model to capture discriminative semantic features, such as object categories in ImageNet-1K \citep{deng2009imagenet}, topic labels in N24News \citep{wang2021n24news}, and hateful content indicators in HatefulMemes \citep{kiela2020hateful}. Overall, this meta-task primarily evaluates the model’s image perception and recognition capabilities.


\paragraph{Visual Question Answering.} In this meta-task, the query consists of an instruction, an image, and a textual question, while the target corresponds to the correct answer. Compared to classification, VQA requires not only accurate visual understanding but also higher-level reasoning that integrates commonsense and world knowledge. For example, in DocVQA \citep{mathew2021docvqa}, the model must interpret visual elements such as charts or curves to extract precise values, whereas in ScienceQA \citep{lu2022scienceqa}, it needs to combine visual recognition with factual knowledge, such as identifying Wyoming as a U.S. state and knowing that Cheyenne is its capital. Overall, this meta-task evaluates the model’s visual understanding, reasoning, and knowledge grounding capabilities.


\paragraph{Information Retrieval.} The retrieval query and target may each consist of text, images, instructions, or their combinations. The objective is to rank candidate items by relevance to the query, where relevance is determined by both semantic correspondence and adherence to the given instruction. For example, VisualNews \citep{liu2020visualnews} tasks the model with retrieving an image caption given a news image, or retrieving the corresponding image given a textual caption. FashionIQ \citep{wu2021fashioniq} and CIRR \citep{liu2021cirr} further evaluate the model’s ability to follow natural-language instructions by retrieving an image that satisfies a specified modification relative to a reference image. Overall, this meta-task primarily assesses the model’s cross-modal alignment and instruction-following capabilities in retrieval. More dataset details are provided in \Cref{dataset_details}.


\begin{table*}[t!]
\centering
\renewcommand{\arraystretch}{1}
\captionsetup{justification=justified,singlelinecheck=false}
\begin{tabularx}{\textwidth}{
    >{\raggedright\arraybackslash}l 
    >{\centering\arraybackslash}X
    >{\centering\arraybackslash}X
    >{\centering\arraybackslash}X
    >{\centering\arraybackslash}X
}
\toprule

\textbf{Dataset} 
& \textbf{\llava}
& \textbf{\qwen}
& \textbf{\mmada}
& \textbf{\lavida} \\
\midrule

VisDial            & \textbf{76.3} & 75.5 & 47.9 & 72.5 \\
CIRR               & 53.6 & \textbf{55.0} & 32.4 & 53.2 \\
VisualNews-t2i     & \textbf{74.4} & 74.1 & 27.9 & 67.2 \\
VisualNews-i2t     & \textbf{78.5} & 74.9 & 32.8 & 62.1 \\
MSCOCO-t2i         & \textbf{71.5} & 71.4 & 53.3 & 70.2 \\
MSCOCO-i2t         & \textbf{68.6} & 68.0 & 50.0 & 55.6 \\
NIGHTS             & \textbf{68.4} & 66.9 & 58.8 & 67.2 \\
WebQA              & 88.8 & \textbf{89.1} & 83.7 & 87.7 \\
\rowcolor{black!6}
FashionIQ          & 19.8 & \textbf{22.2} & 9.2 & 17.4 \\
\rowcolor{black!6}
Wiki-SS-NQ         & \textbf{65.4} & 63.3 & 3.5 & 63.1 \\
\rowcolor{black!6}
OVEN               & 55.5 & \textbf{65.3} & 29.3 & 60.9 \\
\rowcolor{black!6}
EDIS               & \textbf{90.1} & 84.8 & 55.7 & 81.8 \\
\midrule
\textit{All Retrieval} & \textbf{67.6} & 67.5 & 40.4 & 63.2 \\

\bottomrule
\end{tabularx}
\caption{Performance comparison across 12 retrieval tasks. Best results are shown in bold for each task. Tasks highlighted in gray are in-domain, meaning their data have been included during contrastive fine-tuning.}
\label{tab:retrieval_results}
\end{table*}

\subsection{Models}
\label{models}

To compare embeddings from Autoregressive and \diff, we select two representative pretrained models from each paradigm with comparable model sizes. For \auto, we include \llava \citep{llava} and \qwen \citep{qwen2.5-vl}. 
\llava combines a pretrained large language model, \ie Mistral-7B \citep{mistral}, with a pretrained vision encoder, \ie CLIP-L \citep{clip}, aligning visual and textual representations via a projection layer followed by instruction tuning. 
\qwen integrates a redesigned Vision Transformer (ViT) as the vision encoder with the language backbone Qwen2.5-7B \citep{qwen} through a two-layer multi-layer perceptron (MLP).
Empirically, \qwen exhibits stronger multimodal understanding and reasoning capabilities than \llava, achieving 58.6 on the MMMU validation set compared to 35.3 for \llava \citep{mmmu}.


For \diff, we consider two state-of-the-art models: \mmada \citep{mmada} and \lavida \citep{lavida}. \mmada adopts a unified diffusion architecture that employs MAGVIT-v2 \citep{magvit} as the image tokenizer and LLaDA-8B \citep{llada} for text tokenization and sequence modeling. It is trained on a mixture of objectives, including textual reasoning, multimodal understanding, and text-to-image generation. \lavida equips diffusion language models, \eg \dream-7B \citep{dream}, with a vision encoder (SigLIP-400M \citep{siglip2}) and jointly fine-tunes all components for multimodal instruction following. On the MMMU validation set, \lavida achieves a score of 42.6, substantially outperforming \mmada, which attains 30.2. Model details are summarized in \Cref{model_summary}.


\subsection{Implementations}
\label{impl}
\begin{figure*}[t!]
    \centering
    \includegraphics[width=0.95\textwidth]{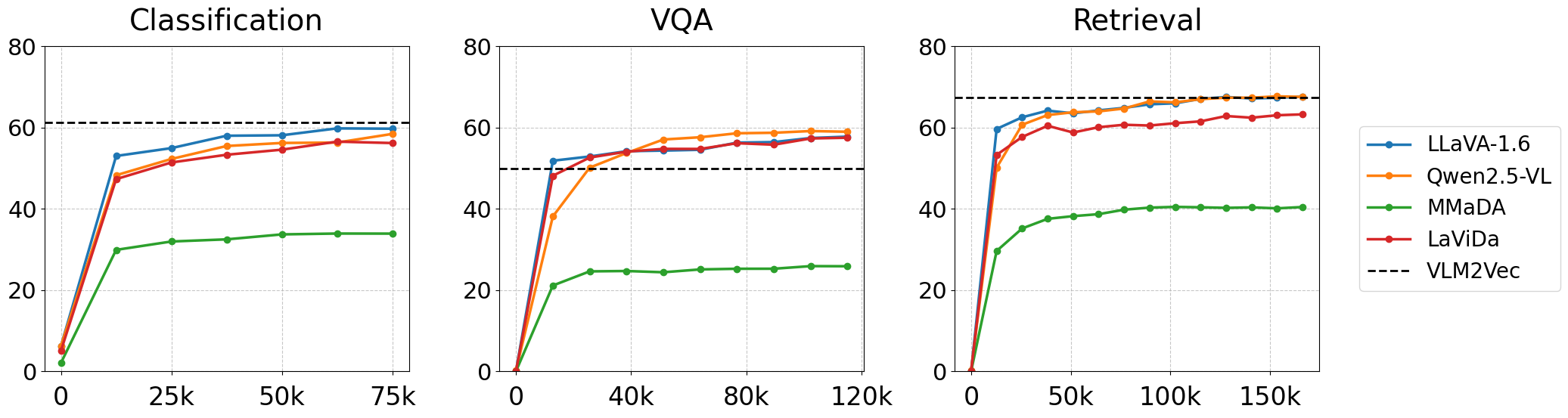}
    \captionsetup{justification=justified, singlelinecheck=false}  
    \caption{
Embedding performance of Autoregressive and Diffusion VLMs under varying fine-tuning data scales.
VLM2Vec reflects performance after large-scale training on 662k samples aggregated across all meta-tasks. In contrast, the fine-tuning curves show that the majority of performance gains are achieved with substantially smaller datasets, with diminishing returns as data scale increases. As VLM2Vec is trained in a multi-task setting, it may underperform on certain tasks (\eg VQA) compared to other models fine-tuned specifically for a single meta-task.
}

    \label{trend}
\end{figure*}

We integrate the selected pretrained models into the VLM2Vec framework \citep{vlm2vec} for fine-tuning and evaluation. We conduct three separate experiments—classification, VQA, and retrieval—each using a distinct set of in-domain and out-of-domain datasets, covering a total of 19 in-domain and 13 out-of-domain datasets.\footnote{Given the poor zero-shot embedding performance observed, we primarily report results after fine-tuning.} Due to resource constraints, we cap each in-domain training dataset at 20k samples, resulting in 76k training samples for classification, 106k for VQA, and 153k for retrieval. Detailed statistics of dataset is provided in \Cref{tab:meta_tasks}. As shown in \Cref{trend}, the fine-tuned models recover most of the performance of VLM2Vec trained at full scale. Additional implementation details are provided in \Cref{impl_details}.

\section{Experiment Results}
\label{results}

\paragraph{Overall, \diff still underperforms compared to \auto, even with bidirectional attention.} As shown in \Cref{tab:classification_results,tab:vqa_results,tab:retrieval_results}, Autoregressive VLM embeddings (i.e., \llava and \qwen) achieve comparable performance across all tasks and consistently outperform \diff embeddings. Compared to the best Autoregressive VLM embeddings, \lavida exhibits a modest performance gap, with average decreases of 3.5 points on classification, 2.5 points on VQA, and 4.4 points on retrieval tasks. A substantially larger performance gap is observed for \mmada embeddings, which is likely due to the limited visual understanding and reasoning capacity of its backbone \citep{mmada}. Notably, \mmada adopts a unified pretraining objective for language modeling and image generation, which might be suboptimal for embedding learning.



\paragraph{While \lavida underperforms on in-domain tasks, it achieves comparable or superior performance on out-of-domain tasks.} For example, on classification tasks, \llava achieves an average score of 76.9 on in-domain tasks and 42.5 on out-of-domain tasks, whereas \lavida attains 67.9 and 44.5, respectively. Although \lavida underperforms \llava by 9.0 points in the in-domain setting, it surpasses \llava by 2.0 points on out-of-domain tasks. Similar trends are observed for VQA and retrieval tasks, implying that \lavida may be more resilient to domain shifts.


\paragraph{Different VLMs exhibit strengths across different categories of VQA tasks.} While \llava embeddings achieve strong performance on most classification and retrieval tasks, VQA performance varies substantially across models. As shown in \Cref{tab:vqa_results}, \llava embeddings excel on tasks requiring world knowledge and commonsense reasoning beyond visual content (\eg OK-VQA and A-OKVQA). In contrast, \qwen performs particularly well on document- and chart-based QA tasks (\eg DocVQA, InfographicsVQA, and ChartQA), reflecting its strength in structured document and chart understanding. \lavida achieves the best performance on compositional reasoning tasks involving objects, attributes, and relations (\eg GQA). These complementary strengths suggest that different VLMs encode distinct inductive biases, making them better suited to different forms of multimodal reasoning.



\paragraph{The text and image embeddings in \diff are not sufficiently aligned, as evidenced by its inferior performance on cross-modal retrieval tasks.} \lavida exhibits a substantial performance gap relative to \auto on image-to-text retrieval tasks such as VisualNews-i2t and MSCOCO-i2t, as shown in \Cref{tab:retrieval_results}. In contrast, it achieves markedly stronger performance on retrieval tasks where both the query and the target involve images, such as CIRR and NIGHTS. This pattern suggests weaker alignment between text and image embeddings in \diff.

\section{Analysis}
\label{analysis}

\paragraph{RQ1: How does the amount of fine-tuning data affect embedding performance?}
\Cref{trend} shows that embedding performance across all models and tasks improves rapidly with only a small amount of data and then quickly saturates, exhibiting clear diminishing returns as data scale increases. Notably, a relatively small fraction of the data already recovers most of the performance achieved by VLM2Vec, indicating that contrastive fine-tuning is highly data-efficient when starting from a strong pretrained checkpoint. Among all models, \llava demonstrates the highest data efficiency across all tasks, suggesting that its pretrained representations already capture well-aligned and semantically rich image–text features.

Despite increased fine-tuning data, the performance gap between Autoregressive and \diff does not close within this regime, underscoring the importance of pretrained checkpoints over data scale alone. This is particularly notable for \mmada: although its image understanding ability is comparable to other models as discussed in Section~\ref{models}, it exhibits a substantial and persistent gap in embedding performance, which cannot be alleviated by additional data. We conjecture that optimization objectives geared toward image generation may negatively impact its suitability for discriminative embedding learning.



\begin{figure}[b!]
    \centering
    \includegraphics[width=0.48\textwidth]{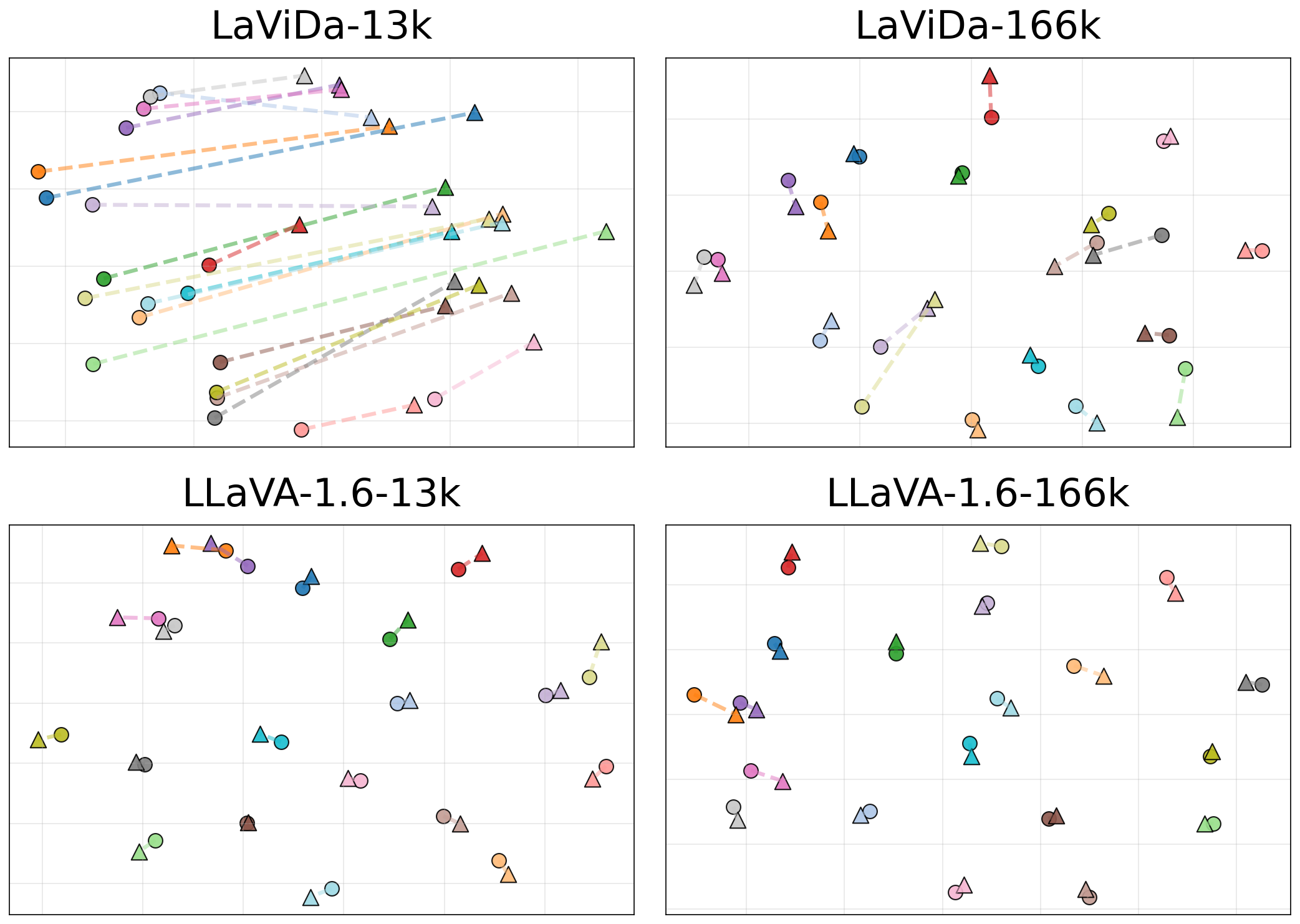}
    \captionsetup{justification=justified, singlelinecheck=false}  
    \caption{t-SNE visualization of query–target embedding pairs on the MSCOCO\_i2t dataset for \lavida and \llava fine-tuned with different amounts of training data. Circles represent query embeddings and triangles represent target embeddings. Dashed lines connect corresponding query–target pairs, indicating their relative distances in the projected embedding space.}
    \label{text_image_alignment}
\end{figure}

\paragraph{RQ2: How well are text and image embeddings aligned?} To qualitatively examine cross-modal alignment, we visualize the query (image) and positive target (text caption) embeddings for 20 examples from the MSCOCO\_i2t task in \Cref{text_image_alignment}. Circles denote image queries and triangles denote their corresponding text targets, with dashed lines indicating paired query–target distances in the embedding space. The 13k training samples represent an early stage of contrastive fine-tuning, while 166k samples correspond to a later stage.

At the early training stage (13k), \lavida exhibits a clear misalignment between image and text embeddings: paired image–text representations are widely separated, and the dashed lines are long and largely parallel, indicating that the model has not yet learned to effectively bridge the modality gap. As fine-tuning progresses to 166k samples, the alignment of \lavida improves, with image and text embeddings moving closer in the shared space. However, noticeable distances between paired embeddings remain, suggesting incomplete cross-modal alignment even at a later training stage.

In contrast, \llava achieves strong image–text alignment already at the early stage of fine-tuning, with most paired embeddings closely co-located and short dashed connections. This alignment remains stable at 166k samples, with consistently smaller query–target distances than those observed for \lavida. Overall, the persistently larger image–text distances for \lavida corroborate its inferior performance on MSCOCO\_i2t, indicating that diffusion-based VLMs are less effectively aligned than autoregressive counterparts under the same contrastive fine-tuning regime.



\paragraph{RQ3: To what extent does fine-tuning the vision encoder influence embedding performance?} 
Different from other models, the vision encoder parameters in \mmada are frozen under the default training strategy, which may limit effective alignment between visual and textual representations. To examine the significance of this design choice, we conduct an additional experiment in which the vision encoder is made trainable during contrastive fine-tuning. However, as shown in \Cref{tab:vqa_results_mmada_frozen_unfrozen}, the resulting performance remains nearly unchanged on average, indicating that fine-tuning the vision encoder has minimal impact on embedding performance. This is likely because the vision encoder contains only about 500M parameters, whereas the majority of representational capacity resides in the $\sim$7B-parameter language backbone, thereby limiting the effect of updating the vision encoder.

\begin{table}[t!]
\centering
\renewcommand{\arraystretch}{1.1}
\setlength{\tabcolsep}{4pt}
\captionsetup{justification=justified,singlelinecheck=false}
\small
\begin{tabularx}{\columnwidth}{
    l
    >{\centering\arraybackslash}X
    >{\centering\arraybackslash}X
}
\toprule
\textbf{Dataset} & \textbf{Frozen} & \textbf{Unfrozen} \\
\midrule
OK-VQA          & \textbf{42.0} & 41.8 \\
A-OKVQA         & 30.4 & \textbf{30.8} \\
DocVQA          & \textbf{17.2} & 16.3 \\
InfographicsVQA & \textbf{17.8} & 17.6 \\
ChartQA         & \textbf{14.3} & 14.1 \\
Visual7W        & \textbf{38.8} & 37.9 \\
\rowcolor{black!6}
ScienceQA       & \textbf{18.9} & 18.2 \\
\rowcolor{black!6}
VizWiz          & 12.9 & \textbf{13.0} \\
\rowcolor{black!6}
GQA             & 51.1 & \textbf{53.6} \\
\rowcolor{black!6}
TextVQA         & 15.2 & \textbf{15.5} \\
\midrule
\textit{All VQA} & 25.9 & 25.9 \\
\bottomrule
\end{tabularx}
\caption{Performance comparison across 10 VQA tasks with frozen versus unfrozen vision encoders for \mmada.}
\label{tab:vqa_results_mmada_frozen_unfrozen}
\end{table}

\section{Conclusion}

This paper presents the first systematic comparative analysis of multimodal embeddings derived from diffusion-based and autoregressive vision language models. Through extensive experiments on classification, visual question answering, and retrieval tasks, we show that Diffusion VLM embeddings underperform their autoregressive counterparts, despite benefiting from bidirectional attention. We attribute this gap primarily to weaker multimodal understanding and less effective cross-modal alignment. We hope our findings provide useful insights for the multimodal embedding community and inform future research on diffusion-based VLMs.

\section*{Limitations}

We have evaluated only two state-of-the-art multimodal diffusion language models, \ie \lavida \citep{lavida} and \mmada \citep{mmada}. Recently, DiffusionVL \citep{diffusionvl} was released. Built upon Qwen2.5-VL rather than a diffusion language model, it achieves a new state-of-the-art performance, suggesting that its embeddings may exhibit strong performance. Contemporarily, \dream-VL \citep{dreamvl} was also released, which extends \dream with multimodal capabilities comparable to top-tier autoregressive VLMs trained on open data across various benchmarks, while exhibiting superior potential in visual planning tasks. We leave the investigation of these newly released models to future work during the paper reviewing process. Due to resource constraints, we limit the fine-tuning scale within 200k samples. Larger-scale experiments with millions of examples could reveal further insights.






\bibliography{custom.bib}

\clearpage
\appendix

\section{Dataset Summary}
\label{dataset_details}

We employ the following datasets, as processed by \citet{vlm2vec}, to evaluate different VLM embeddings. For contrastive fine-tuning, we use xx samples from each training set. For evaluation, we use the full test set.

\subsection{Classification}

\paragraph{ImageNet-1K (I$\to$T)} \citep{deng2009imagenet}
A large-scale object classification dataset with 1K visual categories.

\paragraph{ImageNet-A (I$\to$T)} \cite{hendrycks2021imageneta}
An out-of-distribution variant of ImageNet that evaluates robust object classification on challenging images.

\paragraph{ImageNet-R (I$\to$T)} \cite{hendrycks2021imagenetr}
A robustness benchmark for object classification using artistic and non-photorealistic renderings of ImageNet classes.

\paragraph{VOC2007 (I$\to$T)} \cite{everingham2010pascal}
An object classification benchmark with 20 object categories in real-world images.

\paragraph{N24News (I+T$\to$I)} \cite{wang2021n24news}
A multimodal news classification dataset that assigns image–text pairs to one of 24 news categories.

\paragraph{HatefulMemes (I$\to$T)} \cite{kiela2020hateful}
A multimodal classification task for detecting hateful content in image–text memes.

\paragraph{Places365 (I$\to$T)} \cite{zhou2017places}
A scene classification dataset with 365 semantic environment categories.

\paragraph{SUN397 (I$\to$T)} \cite{xiao2010sun}
A fine-grained scene classification benchmark with 397 scene categories.

\paragraph{ObjectNet (I$\to$T)} \cite{barbu2019objectnet}
A robustness-focused object classification dataset featuring objects under unusual poses and viewpoints.

\paragraph{Country-211 (I$\to$T)} \cite{radford2021clip}
A geolocation classification task that predicts the country in which an image was taken across 211 classes.

\subsection{Visual Question Answering}

\paragraph{OK-VQA (I+T$\to$T)} \cite{marino2019okvqa}
A knowledge-based VQA task that requires external knowledge beyond the image to answer questions.

\paragraph{A-OKVQA (I+T$\to$T)} \cite{schwenk2022aokvqa}
An augmented knowledge-based VQA benchmark emphasizing commonsense and world knowledge reasoning grounded in the image.

\paragraph{DocVQA (I+T$\to$T)} \cite{mathew2021docvqa}
A document VQA task that answers questions over document images with varied layouts and content.

\paragraph{InfographicsVQA (I+T$\to$T)} \cite{mathew2022infographicsvqa}
A document understanding task requiring joint reasoning over text, layout, graphics, and data visualizations in infographics.

\paragraph{ChartQA (I+T$\to$T)} \cite{masry2022chartqa}
A VQA benchmark focused on answering questions about charts through visual and logical reasoning.

\paragraph{ScienceQA (I+T$\to$T)} \cite{lu2022scienceqa}
A multimodal QA task covering elementary science questions grounded in images and textual context.

\paragraph{Visual7W (I+T$\to$T)} \cite{zhu2016visual7w}
A visual grounding and reasoning task that answers \textit{what, where, when, who, why,} and \textit{how} questions about images.

\paragraph{VizWiz (I+T$\to$T)} \cite{gurari2018vizwiz}
A real-world VQA task using images captured by blind users, focusing on answering practical and answerable questions.

\paragraph{TextVQA (I+T$\to$T)} \cite{singh2019textvqa}
A VQA benchmark that requires reading and reasoning over text embedded in images.

\paragraph{GQA (I+T$\to$T)} \cite{hudson2019gqa}
A compositional visual reasoning task that evaluates relational and logical reasoning over real-world images.

\subsection{Retrieval}

\paragraph{VisDial (T$\to$I)} \cite{das2017visdial}
A dialogue-based image retrieval task that retrieves an image given a multi-round question–answer dialogue.

\paragraph{CIRR (I+T$\to$I)} \cite{liu2021cirr}
A composed image retrieval task that retrieves a target image given a reference image and a textual modification.

\paragraph{FashionIQ (I+T$\to$I)} \cite{wu2021fashioniq}
A composed image retrieval benchmark that retrieves a target fashion image given a reference image and a modification description.

\paragraph{VisualNews (I$\to$T, T$\to$I)} \cite{liu2020visualnews}
A cross-modal retrieval task with two setups: retrieving captions given news images (VisualNews-i2t) and retrieving images given captions (VisualNews-t2i).

\paragraph{MSCOCO (I$\to$T, T$\to$I)} \cite{lin2014mscoco}
A cross-modal image–text retrieval benchmark with two setups: retrieving captions from images (MSCOCO-i2t) and retrieving images from captions (MSCOCO-t2i).

\paragraph{WebQA (T$\to$I+T)} \cite{chang2022webqa}
A multimodal retrieval-based QA task that retrieves relevant Wikipedia image–text evidence to answer a question.

\paragraph{NIGHTS (I$\to$I)} \cite{fu2023nights}
A visual similarity retrieval task that retrieves the image most similar to a reference image based on human similarity judgments.

\paragraph{OVEN (I+T$\to$I+T)} \cite{hu2023oven}
A retrieval-based VQA task that retrieves relevant Wikipedia image–text evidence given an image and a visual recognition question.

\paragraph{EDIS (T$\to$I+T)} \cite{liu2023edis}
A cross-modal news retrieval task that retrieves news image–headline pairs given entity-rich text queries.

\paragraph{Wiki-SS-NQ (T$\to$I)} \cite{ma2024wikissnq}
A retrieval-based VQA task that retrieves Wikipedia page screenshots to answer natural language questions.

\begin{table*}[t]
\centering
\renewcommand{\arraystretch}{1.1}
\label{tab:meta_tasks}
\begin{tabularx}{\textwidth}{
    l l c c c c c c
}
\toprule
\textbf{Meta-Task} & \textbf{Dataset} & \textbf{Query} & \textbf{Target} & \textbf{OOD?} & \textbf{\#Training} & \textbf{\#Eval} & \textbf{\#Candidates} \\
\midrule

\multirow{10}{*}{Classification} 
& ImageNet-1K   & I     & T &   & 20K & 1000 & 1000 \\
& N24News       & I+T   & I &   & 20K & 1000 & 24   \\
& HatefulMemes  & I     & T &   & 8K  & 1000 & 2    \\
& VOC2007       & I     & T &   & 8K  & 1000 & 20   \\
& SUN397        & I     & T &   & 20K & 1000 & 397  \\
& Place365      & I     & T & \checkmark & - & 1000 & 365 \\
& ImageNet-A    & I     & T & \checkmark & - & 1000 & 1000 \\
& ImageNet-R    & I     & T & \checkmark & - & 1000 & 200  \\
& ObjectNet     & I     & T & \checkmark & - & 1000 & 313  \\
& Country-211   & I     & T & \checkmark & - & 1000 & 211  \\

\midrule

\multirow{10}{*}{VQA}
& OK-VQA            & I+T & T &   & 9K  & 1000 & 1000 \\
& A-OKVQA           & I+T & T &   & 17K & 1000 & 1000 \\
& DocVQA            & I+T & T &   & 20K & 1000 & 1000 \\
& InfographicVQA    & I+T & T &   & 20K & 1000 & 1000 \\
& ChartQA           & I+T & T &   & 20K & 1000 & 1000 \\
& Visual7W          & I+T & T &   & 20K & 1000 & 1000 \\
& ScienceQA         & I+T & T & \checkmark & - & 1000 & 1000 \\
& VizWiz            & I+T & T & \checkmark & - & 1000 & 1000 \\
& GQA               & I+T & T & \checkmark & - & 1000 & 1000 \\
& TextVQA           & I+T & T & \checkmark & - & 1000 & 1000 \\

\midrule

\multirow{12}{*}{Retrieval}
& VisDial            & T   & I     &   & 20K & 1000 & 1000 \\
& CIRR               & I+T & I     &   & 20K & 1000 & 1000 \\
& VisualNews.t2i     & T   & I     &   & 20K & 1000 & 1000 \\
& VisualNews.i2t     & I   & T     &   & 20K & 1000 & 1000 \\
& MSCOCO.t2i         & T   & I     &   & 20K & 1000 & 1000 \\
& MSCOCO.i2t         & I   & T     &   & 20K & 1000 & 1000 \\
& NIGHTS             & I   & I     &   & 16K & 1000 & 1000 \\
& WebQA              & T   & I+T   &   & 17K & 1000 & 1000 \\
& OVEN               & I+T & I+T   & \checkmark & - & 1000 & 1000 \\
& FashionIQ          & I+T & I     & \checkmark & - & 1000 & 1000 \\
& EDIS               & T   & I+T   & \checkmark & - & 1000 & 1000 \\
& Wiki-SS-NQ         & T   & I     & \checkmark & - & 1000 & 1000 \\

\bottomrule
\end{tabularx}
\caption{Summary of training and evaluation tasks across classification, VQA, and retrieval. I and T denote image and text modalities, respectively. OOD indicates out-of-distribution evaluation. Training set sizes larger than 20K are capped at 20K.}
\end{table*}

\section{Model Summary}
\label{model_summary}

The baseline model details are presented in \Cref{tab:vlm_arch_comparison}.

\begin{table*}[t]
\centering
\renewcommand{\arraystretch}{1.15}
\setlength{\tabcolsep}{6pt}

\begin{tabularx}{\textwidth}{
    >{\raggedright\arraybackslash}l
    >{\raggedright\arraybackslash\hsize=0.65\hsize}X
    >{\raggedright\arraybackslash\hsize=1.55\hsize}X
    >{\raggedright\arraybackslash\hsize=0.65\hsize}X
}
\toprule
\textbf{Model} & \textbf{Vision encoder} & \textbf{Vision encoder Update Strategy} & \textbf{Backbone LLM} \\
\midrule

Qwen2.5-VL
& Redesigned ViT
& Vision encoder trained from scratch.\par
  Stage 1: Vision encoder is trained (other modules frozen).\par
  Stage 2: Vision encoder + LLM + adapters are jointly trained.\par
  Stage 3: Instruction fine-tuning with supervised multimodal data.
& Qwen2.5-7B \\
\midrule

LLaVA-1.6
& CLIP-L
& Visual encoder frozen.\par
  Stage 1: Only the projection matrix is updated.\par
  Stage 2: The projection matrix and the LLM are updated.
& Mistral-7B \\
\midrule

MMaDA
& MAGVIT-v2
& Vision encoder frozen.
& LLaDA-8B \\
\midrule

LaViDa
& SigLIP-400M
& Stage 1: Only the projection network is updated (vision encoder frozen).\par
  Stage 2: All parts including the vision encoder are trained end-to-end.
& Dream-7B \\
\bottomrule
\end{tabularx}

\caption{Comparison of VLM architectures, including vision encoders, trainable visual components, and text backbones.}
\label{tab:vlm_arch_comparison}
\end{table*}

\section{Implementation Details}
\label{impl_details}

For all the experiments, the temperature parameter in the loss function is set to 0.02, with a batch size of 256. For LoRA fine-tuning, we use a rank of 8. All models are trained for approximately one epoch, corresponding to 300 steps for classification, 450 steps for VQA, and 650 steps for retrieval. For Autoregressive VLMs, we take the EOS token representation as the embedding vector. But for \diff, we compute the embedding by averaging the hidden states across the entire sequence, reflecting the bidirectional nature of \diff. All experiments are conducted on two NVIDIA H200 GPUs. Following \citet{vlm2vec}, we use Precision@1 as the main evaluation metric, which measures the percentage of queries for which the top-ranked candidate matches the ground-truth target.

\end{document}